\documentclass[prd,aps,preprint,showpacs,nofootinbib]{revtex4}
\usepackage{epsfig}
\usepackage{amsmath}

\begin{document}

\preprint{RBRC-718}\preprint{LBNL-63750}

\title{Collins Asymmetry at Hadron Colliders}

\author{Feng Yuan}
\email{fyuan@lbl.gov} \affiliation{Nuclear Science Division,
Lawrence Berkeley National Laboratory, Berkeley, CA
94720}\affiliation{RIKEN BNL Research Center, Building 510A,
Brookhaven National Laboratory, Upton, NY 11973}

\begin{abstract}
We study the Collins effect in the azimuthal asymmetric
distribution of hadrons inside a high energy jet in the single
transverse polarized proton proton scattering. From
the detailed analysis of one-gluon and two-gluon exchange
diagrams contributions, the Collins function is found the same as
that in the semi-inclusive deep inelastic scattering and $e^+e^-$
annihilations. The eikonal propagators in these diagrams do
not contribute to the
phase needed for the Collins-type single spin asymmetry, and the
universality is derived as a result of the Ward identity. We
argue that this conclusion depends on the momentum flow
of the exchanged gluon and the kinematic constraints in the
fragmentation process, and is generic and model-independent.

\end{abstract}

\maketitle

\newcommand{\be}{\begin{equation}}
\newcommand{\ee}{\end{equation}}
\newcommand{\ben}{\[}
\newcommand{\een}{\]}
\newcommand{\beqn}{\begin{eqnarray}}
\newcommand{\eeqn}{\end{eqnarray}}
\newcommand{\Tr}{{\rm Tr} }

\section{Introduction}

Single-transverse spin asymmetries (SSA) in hadronic processes have a long history \cite{E704,Bunce}.
Recent experimental measurements of SSAs in polarized semi-inclusive lepton-nucleon
deep inelastic scattering (SIDIS) \cite{hermes,dis}, in hadronic
collisions \cite{star,phenix,brahms}, and in the relevant
$e^+e^-$ annihilation process \cite{belle}, have renewed the
theoretical interest in SSAs and in understanding their roles in
hadron structure and Quantum Chromodynamics (QCD).
There are several approaches to understanding
SSAs within the QCD framework \cite{review,Efremov,qiu}.
Transverse-momentum-dependent
(TMD) parton distributions and fragmentation functions, and their
relevance for semi-inclusive DIS, the Drell-Yan process,
di-hadron production in $e^+e^-$ annihilations, and the
single-inclusive hadron production at hadron colliders have been
investigated \cite{RalSop79,ColSop81,ColSop81p,ColSopSte85, Siv90,Col93,Ans94,
MulTan96,BroHwaSch02,Col02,BelJiYua02,BoeMulPij03,Boer:1997mf,Efremov:2006qm,{Anselmino:2007fs}}.

Two important contributions from these TMD parton distributions
and fragmentation functions have been mostly discussed in the
last few years: the Sivers quark distribution and the Collins
fragmentation function. The Sivers quark distribution \cite{Siv90}
represents a distribution of unpolarized quarks in a transversely
polarized nucleon, through a correlation between the quark's
transverse momentum and the nucleon polarization
vector. The existence of the Sivers function
requires final/initial-state interactions~\cite{BroHwaSch02}, and an interference
between different helicity Fock states of the nucleon~\cite{BroHwaSch02,{Ji:2002xn}}.
The Collins function represents a correlation between the transverse
spin of the fragmenting quark and the transverse momentum of the
hadron relative to the ``jet axis'' in the fragmentation process.
Like the Sivers function, it vanishes when integrated over all
transverse momentum.

One of the most nontrivial properties associated with the Sivers
and Collins functions are their universality properties. Although
they both belong to the so-called ``naive-time-reversal-odd" functions, they
do have different universality properties. For the quark Sivers
function, because of the initial/final state interaction
difference, they differ by signs for the SIDIS and Drell-Yan
processes~\cite{BroHwaSch02,Col02,{Ji:2006ub}}.
This non-universality has also been extended to other processes,
such as the dijet-correlation in hadronic reactions, where it was
found that both initial and final state interactions contribute to
the SSA, and there exists non-trivial relation between this
and those in the SIDIS and Drell-Yan processes \cite{gl,QVY,CQ},
and a standard TMD factorization breaks down~\cite{CQ}.

On the other hand, there have been several studies showing that
the Collins function is universal between different processes,
primarily in the SIDIS and $e^+e^-$ annihilation
\cite{Metz:2002iz,Collins:2004nx,Amrath:2005gv}. In these
discussions, the gauge links in the fragmentation functions do
not play a crucial role to leading to a nonzero Collins function
though they are important to retain the gauge invariance, whereas
it has been well understood that the gauge links in the parton
distributions play very important roles to obtain non-zero
quark Sivers function.

The Collins effect in the fragmentation process and its universality
has been recently extended to the hadron production in $pp$ collisions
\cite{collins-s}, where the azimuthal distribution of hadrons inside a high
energy jet can probe the Collins fragmentation function and the
quark transversity distribution \cite{Jaffe:1991kp} in the single
transverse polarized nucleon-nucleon scattering.
In this paper, we will give the detailed derivation of these
results, and argue that the
universality is in general and model-independent.

We are interested in the hadron production from the fragmentation
of a transversely polarized quark which inherit transverse spin
from the incident nucleon through transverse spin transfer in the
hard partonic scattering processes
\cite{Stratmann:1992gu,{Collins:1993kq},umbeto}.  As shown in Fig.~1,
we study the process,
\begin{equation}
p(P_A,S_\perp)+p(P_B)\to jet (P_J)+X\to H(P_h)+X\ , \label{e1}
\end{equation}
where a transversely polarized proton with momentum $P_A$ scatters
on another proton with momentum $P_B$, and produces a jet with
momentum $P_J$ (transverse momentum $P_\perp$ and rapidity $y_1$
in the Lab frame). The three momenta of $P_A$, $P_B$ and $P_J$
form the so-called reaction plane. Inside the produced jet,
the hadrons are distributed around the jet axes.
A particular hadron $H$ will carry certain longitudinal momentum
fraction $z_h$ of the jet, and its transverse momentum $P_{hT}$ relative
to the jet axis will define an azimuthal angle with the reaction
plane: $\phi_h$, shown in Fig.~1. Thus, the hadron's momentum
is defined as $P_h=z_h P_J+P_{hT}$. The relative transverse momentum
$P_{hT}$ is orthogonal to the jet's momentum $P_J$: $\vec{P}_{hT}\cdot \vec{P}_J=0$.
Similarly, we can define the
azimuthal angle of the transverse polarization vector of the
incident polarized proton: $\phi_s$.

\begin{figure}[t]
\begin{center}
\includegraphics[width=9cm]{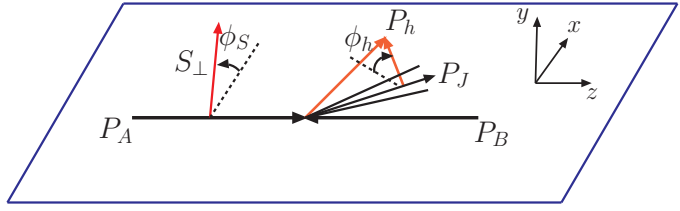}
\end{center}
\vskip -0.4cm \caption{\it Illustration of the kinematics for the
azimuthal distribution of hadrons inside a jet in $pp$
scattering.} \label{fig1}
\end{figure}

The leading order contribution to the jet production in $pp$
collision comes from $2\to 2$ sub-processes, where two jets are
produced back-to-back in the transverse plane. For the reaction
process of (1), one of the two jets shall fragment into the final
observed hadron. In this paper, we study the physics in the
kinematic region of $P_{hT}\ll P_\perp$.
The unpolarized cross section contribution from the
partonic $2\to 2$ process $ab\to qc$ where the final state quark
$q$ fragments into final observed hadron $H$, can be written as
\begin{eqnarray}
\frac{d\sigma^{uu}}{dy_1dy_2dP_\perp^2dzd^2P_{hT}}=\frac{d\sigma^{uu}}{d{\cal
P.S.}}=\sum_{b=q,g}x'f_b(x') x f_a(x)D_q(z_h,P_{hT})\times
H_{ab\to qc}^{\rm uu}\ ,
\end{eqnarray}
where $d{\cal P.S.}={dy_1dy_2dP_\perp^2dzd^2P_{hT}}$ represents
the phase space for this process, $y_1$ and $y_2$ are rapidities
for the jet $P_J$ and the balancing jet, respectively, $P_\perp$
is the jet transverse momentum, and the final observed hadron's
kinematic variables $z_h$ and $P_{hT}$ are defined above. Here,
$x$ and $x'$ are the momentum fractions carried by the parton
``$a$" and``$b$" from the incident hadrons, respectively. In the above
equation, $f_a$ and $f_b$ are the associated parton distributions,
and $D_q(z_h,P_{hT})$ is the TMD quark fragmentation function.
The hard factors $H_{ab\to qc}$ are equal to the
partonic differential cross section for the relevant subprocess:
$H_{ab\to qc}=d\hat\sigma/d\hat t|_{ab\to qc}$. Similarly,
the differential cross section for the transverse-spin
dependent scattering process can be written as
\begin{eqnarray}
\frac{d\sigma(S_\perp)}{d{\cal P.S.}}=\sum_{b=q,g}x'f_b(x')
x\delta q_T(x)\delta \hat q(z_h,P_{hT})\frac{\epsilon^{\alpha\beta}S_\perp^\alpha}{M_h}\nonumber\\
\times \left[P_{hT}^\beta-\frac{P_B\cdot P_{hT}}{P_B\cdot
P_J}P_{J}^\beta\right]\times H_{qb\to qb}^{\rm Collins}\ ,
\end{eqnarray}
$\epsilon_\perp^{\alpha\beta}=\epsilon^{\mu\nu\alpha\beta}P_{A\mu}P_{B\nu}/P_A\cdot
P_B$ with convention $\epsilon^{0123}=1$, and $H_{qb\to qb}^{\rm
Collins}$ is the hard factor for the partonic channel $qb\to qb$.
Here, $\delta q_T(x)$ (also noted by $\delta q$, $h_{1q}$ and $\Delta_T q$ in
the literature) is the quark transversity distribution, and $\delta
\hat q$ the Collins fragmentation function \cite{Col93} (also
noted as $\Delta \hat D$ or $H_1^\perp$ in the literature).

It was argued that the
Collins function is universal between the above process and other
processes such as $e^+e^-$ annihilation and SIDIS~\cite{collins-s}.
As an example, we will demonstrate this universality for the
particular partonic channel $qq'\to qq'$ contribution to our
process, and all other channels will follow accordingly. For
convenience, we list the hard factors for this channel,
\begin{eqnarray}
H_{qq'\to qq'}^{\rm uu}=\frac{\alpha_s^2\pi}{\hat
s^2}\frac{N_c^2-1}{4N_c^2}\frac{2(\hat s^2+\hat u^2)}{-\hat
t^2},~~~H_{qq'\to qq'}^{\rm Collins}=\frac{\alpha_s^2\pi}{\hat
s^2}\frac{N_c^2-1}{4N_c^2}\frac{4\hat s\hat u}{-\hat t^2} \
,\label{e8}
\end{eqnarray}
for the unpolarized and single-transverse-spin polarized cross
sections, respectively. Here $\hat s$, $\hat t$, and $\hat u$ are
the usual partonic Mandelstam variables.

The rest of this paper is organized as follows. In Sec.II, we calculate
the differential cross sections for the unpolarized and single-spin
dependent scattering processes from $qq'\to qq'$ channel contributions,
and demonstrate the universality of the Collins function. Especially,
we will present a detailed calculation for one-gluon exchange
diagrams which are essential for the universality argument.
An extension to two-gluon exchange diagrams is
presented in Sec.III. We summarize our paper in Sec.IV.

\section{Universality of the Collins function}

We follow the model used in Ref.~\cite{Col93} to calculate
the quark fragmentation into a pion. As shown in Fig.~2(a), a quark
(with momentum $k$) fragments into a pion (with momentum $P_h$) by
the vertex from a model described in \cite{Manohar:1983md}. A simple
calculation will give the unpolarized quark fragmentation function
\cite{Col93},
\begin{equation}
D_q(z_h,P_{hT})=\frac{g^2}{16\pi^3}\frac{z_h^2}{\vec{P}_{hT}^2+z_h^2M^2}
\ ,\label{unf}
\end{equation}
where $g$ is the coupling between the quark and pion, $M$ is the
quark mass.

\begin{figure}[t]
\begin{center}
\includegraphics[width=9cm]{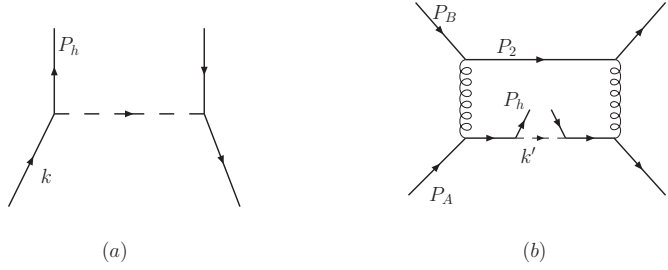}
\end{center}
\vskip -0.4cm \caption{\it Quark fragmentation to pion production
(a), and in $pp$ scattering in a model described in \cite{Manohar:1983md} (b).}
\end{figure}

We can also use this model to calculate pion production in
hadronic process of (1). In Fig.~2(b), we show the Feynman diagram
for the typical partonic channel $qq'\to qq'$ contribution, where
the initial quarks have momenta $P_A$ and $P_B$, respectively. In
the final state, the produced pion has momentum $P_h$, and the
associated final state quark has momentum $k'$, whereas the
balancing jet has momentum $P_2$. We further introduce a
light-like momentum $\hat k$: $\hat k^2=0$, which represents the
dominant component of the fragmenting quark's momentum. It can be
parameterized as follows,
\begin{equation}
\hat k=-\frac{\hat u}{\hat s} P_A-\frac{\hat t}{\hat s}
P_B+\vec{P}_\perp \ ,
\end{equation}
where $P_\perp$ is the transverse momentum for the fragmenting
quark in the Lab frame, $\hat s$, $\hat t$ and $\hat u$ as
mentioned above, are the usual partonic Madelstam variables for
this partonic process: $\hat s=2P_A\cdot P_B$, $\hat t=-2P_A\cdot
\hat k$, and $\hat u=-2P_B\cdot \hat k$. In our discussions, the
jet's transverse momentum $P_\perp$ (in the lab frame) is the
large momentum scale at the same order as $\hat s$,
$\hat t$ and $\hat u$. Of course, the full momentum of the
fragmenting quark $P_1=P_h+k'$ is off-shell in this diagram.
However, its off-shellness is much smaller than
 $P_\perp$. In order to formulate the final state hadron's
momentum, we introduce a conjugate light-like vector $\hat n$:
$\hat n^0=\hat k^0$ and $\vec{\hat n}=-\vec{\hat k}$. It is
convenient to define this momentum in the center of mass frame of
the two incident momenta $P_A$ and $P_B$. In this frame, we have
\begin{equation}
\hat n=P_2=-\frac{\hat t}{\hat s} P_A-\frac{\hat u}{\hat s}
P_B-\vec{P}_\perp \ ,
\end{equation}
which happens to be the momentum of the balancing jet. From above,
we have $\hat k^2=\hat n^2=0$ and $\hat k\cdot \hat n=\hat s/2$.
In the following calculations, we will work in this particular
frame. We emphasize that our results do not depend on the
frame.

With the above two momenta, we can formula the final state pion's
momentum as
\begin{equation}
P_h=z_h\hat k+\frac{\vec{P}_{hT}^2}{2z_h\hat k\cdot\hat n}\hat
n+\vec{P}_{hT}\ ,
\end{equation}
where $z_h=P_h\cdot \hat n/\hat k\cdot \hat n$ is the momentum
fraction of the fragmenting quark carried by the pion in the
final state, $P_{hT}$ is the transverse momentum relative to the
fragmenting quark momentum $\hat k$: $P_{hT}\cdot \hat k=0$ and
$P_{hT}\cdot\hat n=0$. In the above parameterization, we have neglect the
pion mass, which is not relevant in our calculations. Similarly,
we can formulate the associated final state quark momentum $k'$
as,
\begin{equation}
k'=(1-z_h)\hat k+\frac{\vec{P}_{hT}^2+M^2}{2(1-z_h)\hat k\cdot\hat
n}\hat n-\vec{P}_{hT}\ ,
\end{equation}
where we have kept the quark mass, because it will be relevant
for the nonzero single spin asymmetry discussed below.

With the above decompositions for the relevant momenta, it is
straightforward to calculate the Feynman diagrams for this
process in Fig.~2(b). In the calculations, we will utilize the
power counting method to keep the leading order contributions,
and neglect all higher order corrections of
$P_{hT}/P_\perp$ or $M/P_\perp$. By doing that, we can separate
the short distance physics (at the scale of $P_\perp$) from the
long distance physics (at the scale of $P_{hT}$ and $M$).

Finally, the cross section contribution from Fig.~2(b) will be,
\begin{equation}
\frac{d\sigma^{uu}}{d{\cal P.S.}}=\frac{\alpha_s^2\pi}{\hat
s^2}\frac{N_c^2-1}{4N_c^2}\frac{2(\hat s^2+\hat u^2)}{\hat
t^2}\frac{g^2}{16\pi^2}\frac{z_h^2}{P_{hT}^2+z_h^2M^2} \ ,
\end{equation}
in the limit of $P_{hT}\ll P_\perp$. This result is indeed
factorized into the hard factor $H_{qq'\to qq'}^{\rm uu}$ in
Eq.~(\ref{e8}) times the fragmentation function in Eq.~(\ref{unf})
calculated from Fig.~2(a).

\begin{figure}[t]
\begin{center}
\includegraphics[width=9cm]{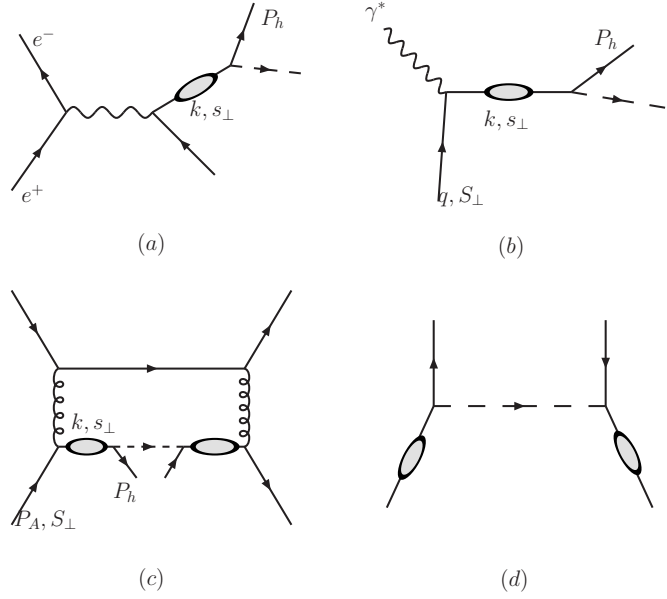}
\end{center}
\vskip -0.4cm \caption{\it Universality of the Collins function
in $e^+e^-$ (a), deep inelastic scattering (b), and $pp$
scattering (c), when we have dressed quark propagator associated
with the fragmenting quark in these processes. The universal
Collins function can be calculated from the diagram in (d). The
blobs in the diagrams represent the dressed quark propagator in
this model. }
\end{figure}

Now, we turn to discuss the SSA in this process. We need to
generate a phase from the scattering amplitudes to have a
non-vanishing SSA. As suggested in \cite{Col93}, the dressed
quark propagator in this model may contribute to such a
phase. Similarly, the vertex correction to the quark-pion vertex
can also contribute a phase \cite{{Bacchetta:2001di}}. If the phase
comes from the above sources, it is easy to argue the
universality of the Collins function between our process and the
SIDIS/$e^+e^-$ process, because they are the same. For example,
as we show in Fig.~3, the dressed quark propagator associated
with the fragmenting quark can contribute to a nonzero phase
\cite{Col93}, which will contribute the same to the Collins
function in all these three processes: $e^+e^-$ annihilation,
SIDIS, and hadron production in $pp$ scattering. This propagator
can be parameterized as: $i(A\!\not P_1+BM)/(P_1^2-M^2)$
\cite{Col93}, where $A$ and $B$ are complex numbers. Following the
above calculations for the unpolarized cross section, we will
find the single-spin dependent cross section for process (1) from
Fig.~3(c) can be written as
\begin{equation}
\frac{d\sigma(S_\perp)}{d{\cal
P.S.}}={\epsilon^{\alpha\beta}S_\perp^\alpha}
\left[P_{hT}^\beta-\frac{P_B\cdot P_{hT}}{P_B\cdot
P_J}P_{J}^\beta\right]\frac{\alpha_s^2\pi}{\hat s^2}\frac{N_c^2-1}{4N_c^2}\frac{4\hat s\hat u}{-\hat t^2}
\frac{g^2}{16\pi^2}\frac{z_h^2(1-z_h)2M{\rm
Im}(A^*B)}{(P_{hT}^2+z_h^2M^2)^2} \ ,
\end{equation}
where again we only keep the leading order contribution in the
limit of $P_{hT}\ll P_\perp$ and $M\ll P_\perp$. In the derivation
of the above result, the following identity has been used to
simplify the final expression,
\begin{equation}
\epsilon^{\alpha\beta}\left[\hat sP_\perp\cdot S_\perp
P_\perp^\alpha P_{hT}^\beta+(\hat u-\hat t)P_B\cdot
P_{hT}P_\perp^\alpha S_\perp^\beta-\hat t\hat uP_{hT}^\alpha
S_\perp^\beta\right] =0\ ,
\end{equation}
which holds in our working frame. The above differential cross
section can be factorized into the Collins function calculated
from the dressed quark propagator from Fig.~3(d) \cite{Col93} and
the hard factor $H_{qq'\to qq'}^{\rm Collins}$ from Eq.~(\ref{e8})
in this partonic channel $qq'\to qq'$, and this Collins function
will be the same as that in $e^+e^-$ and SIDIS processes in
Fig.~3(a) and 3(b).

Similarly, the vertex corrections contributions to the Collins
function can be analyzed accordingly, and the same factorization
and universality of the Collins function will follow.

\begin{figure}[t]
\begin{center}
\includegraphics[width=10cm]{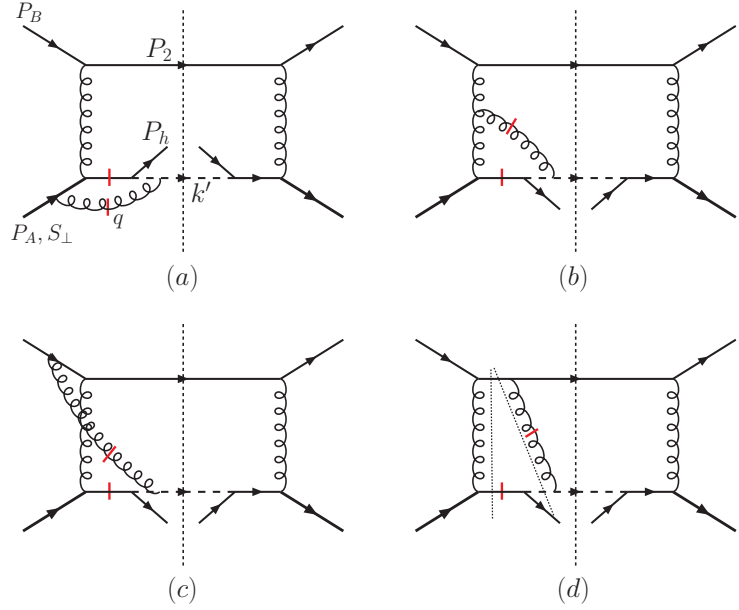}
\end{center}
\vskip -0.4cm \caption{\it Gluon exchange diagrams contributions
to the Collins asymmetry in $pp$ collisions. The short bars
indicate the pole contributions to the phase needed for a
non-vanishing SSA. The additional two cuts in (d) cancel out each
other.} \label{fig2}
\end{figure}

The main issue of the universality discussion concerns the extra
gluon exchange contribution between the spectator and hard
partonic part \cite{Metz:2002iz}. For example, in our case,
because the hadron is colorless while the quark is colored, the
remanet in the fragmentation process will be also colored. Thus
the gluon exchanges between the remanet and the other parts of
the scattering amplitudes become essential. In Fig.~4, we have
shown all these interactions, including the gluon attachments to
the incident quarks (a,c), and final state balancing quark (d)
and the internal gluon propagator (b). These diagrams are much
more complicated than those discussed in \cite{Metz:2002iz} for
SIDIS and $e^+e^-$ processes, where there is only one diagram
contribution in both cases. Therefore, the universality argument
for the Collins function is not straightforward. However,
the dominant contribution to the fragmentation function comes from
the kinematic region where the exchanged gluon is parallel to the
final state hadron \cite{css-fac}. Otherwise, their contributions
will be power suppressed in the limit of $P_{hT}\ll P_\perp$ or belong to
a soft factor.
For these collinear gluon interactions, we can use eikonal approximation and
Ward identity to sum them together to form the gauge
link in the definition of the fragmentation function \cite{css-fac}.

Meanwhile, we notice that the contributing phases of the diagrams
in Fig.~4 come from the cuts through the internal propagators in
the partonic scattering amplitudes
\cite{{BroHwaSch02},{Metz:2002iz}}. In Fig.~4, we labeled these
cut-poles by short bars in the diagrams. From our calculations,
we find that all these poles come from a cut through the
exchanged gluon and the fragmenting quark in each diagram, and
all other contributions either vanish in the leading order
contribution or cancel out each other. For example, in Fig.~2(d),
we show two additional cuts, which contribute however opposite to
each other and cancel out completely. To see this cancellation
more clearly, we can write down the momentum integral of the
exchange-gluon,
\begin{eqnarray}
\int\frac{d^4q}{(4\pi)^4}{\cal M}(q)
\frac{1}{(k'-q)^2-M^2+i\epsilon}\frac{1}{q^2+i\epsilon}
\frac{1}{(P_2+q)^2+i\epsilon}\nonumber\\
\times \frac{1}{(P_2-P_B+q)^2+i\epsilon}
\frac{1}{(P_h+k'-q)^2-M^2+i\epsilon} \ ,
\end{eqnarray}
where ${\cal M}(q)$ represents the denominators coming from the
scattering amplitude. By power counting analysis, the dominant
contribution to the fragmentation comes from the kinematic region
of $q$ being parallel to the final state hadron's momentum $q\sim
P_h$. From this fact, we can parameterize $q$ in terms of $\hat
k$ and $\hat n$, and define $q^+=q\cdot \hat n/\hat k\cdot \hat
n$ and $q^-=q\cdot \hat k/\hat k\cdot \hat n$. Thus the integral
of momentum $q$ becomes $d^4q=\hat k\cdot \hat n d
q^+dq^-d^2q_{T}$, where $q_T$ is the transverse momentum relative
to the jet momentum $\hat k$. Because $q$ is parallel to $\hat
k$, $q^+$ will be order 1, whereas $q^-$ will be order of
$q_{T}^2/q^+$. When we perform the integrals of $dq^+dq^-$, we
need to take two poles from the above propagators to obtain a
nonzero Collins asymmetry. These poles will
form a cut through the Fenyman diagram. Physically, these cuts
represent the kinematic allowed final state re-scattering in the
diagram.

By examining the behaviors of the propagators in the above
kinematic region, we further notice that the $t-$channel gluon
propagator $1/(P_2-P_B+q)^2$ does not contribute to a pole. This
is because this propagator is far off-shell: $(P_2-P_B)^2=\hat
t\sim -|\vec{P}_\perp|^2$. If we take a pole from this propagator,
we have to constrain the momentum of $q$ being proportional to
$P_2$ and $P_B$, whose contribution will be power suppressed.
Thus, we shall calculate the pole contributions from other
propagators. In Fig.~4{d}, we show three possible cuts which are
kinematic allowed for this diagram. Two of them are associated
with the propagator $1/(P_2+q)^2$. This propagator involves large
momentum $P_2$, and can be simplified by using the eikonal
approximation,
\begin{equation}
\frac{1}{(P_2+q)^2+i\epsilon}\approx \frac{1}{2P_2\cdot \hat
k}\frac{1}{q^++i\epsilon} \ .
\end{equation}
The pole contribution from this propagator is proportional to
$\delta (q^+)$. With this delta function, the integral over $q^-$
vanishes, because the rest poles are in the same half plane of
$q^-$,
\begin{eqnarray}
&&\int\frac{dq^-}{2\pi}\frac{1}{(k'-q)^2-M^2+i\epsilon}\frac{1}{(P_h+k'-q)^2-M^2+
i\epsilon}\cdots
\nonumber\\
&&\sim\int\frac{dq^-}{2\pi}\frac{1}{2(k^{\prime +}-q^+) q^-
+\cdots +i\epsilon}\frac{1}{2(P_h^++k^{\prime+}-q^+)q^-+\cdots
+i\epsilon}\cdots =0 \ ,
\end{eqnarray}
where we have used the fact that $k^{\prime +}-q^+>0$ and
$P_h^++k^{\prime +}-q^+>0$. This means that the two cuts
associated with the propagator $1/(P_2+q)^2$ cancel out each
other. The above result depends on the momentum flow of $q$ in
this diagram and the time-like process in the fragmentation region
requiring that $k^{\prime +}>0$ and $P_h^+>0$.

Therefore, the only contribution to the nonzero SSA associated
with the Collins effect comes from the cut going
through the fragmenting quark and the exchange-gluon, as we
labeled by short bars in this diagram. Summarizing the above
analysis, we find that the contribution from this diagram can be
written as
\begin{eqnarray}
\frac{\hat k\cdot \hat
n}{(P_2-k_1)^2(P_2-P_B)^2}\int\frac{dq^+dq^-d^2q_T}{(2\pi)^4}\frac{1}{q^+}
\frac{1}{1-q^+}\frac{1}{(k'-q)^2-M^2}{\cal M}(q)\nonumber\\
\times\delta(q^2)\delta\left((P_h+k'-q)^2-M^2\right) \ ,
\label{eikonal}
\end{eqnarray}
where we have made the eikonal approximation for the propagators
$1/(P_2+q)^2$ and $1/(P_2-P_B+q)^2$.

Similar analysis can be done for all other diagrams, and we find
that their contributions come from the same poles of the
fragmenting quark and the exchange-gluon. Therefore, their
contributions will have the similar expression as
Eq.~(\ref{eikonal}) with the same delta functions in the integral:
$\delta(q^2)\delta\left((P_h+k'-q)^2-M^2\right)$ and the
propagator $1/((k'-q)^2-M^2)$. Thus the contributions from all
these diagrams can be summed together. In this sum, we notice
that the different diagrams have different color-factors,
\begin{eqnarray}
{\rm 4(a)}:&& \frac{1}{N_c^2}{\rm Tr}[T^aT^cT^bT^c]\times {\rm
Tr}[T^aT^b]=C_F\times
\frac{N_c^2-1}{4N_c^2}+\frac{1}{N_c^2}\frac{if_{abc}}{2}{\rm
Tr}[T^aT^bT^c]
\ ,\nonumber\\
{\rm 4(b)}:&& \frac{1}{N_c^2}{\rm Tr}[T^aT^cT^b]\times {\rm
Tr}[T^aT^d]if_{dbc}=-\frac{1}{N_c^2}\frac{if_{abc}}{2}{\rm
Tr}[T^aT^bT^c]
\ ,\nonumber\\
{\rm 4(c)}:&& \frac{1}{N_c^2}{\rm Tr}[T^aT^cT^b]\times {\rm
Tr}[T^aT^bT^c]
\ ,\nonumber\\
{\rm 4(d)}:&& \frac{1}{N_c^2}{\rm Tr}[T^aT^cT^b]\times {\rm
Tr}[T^aT^cT^b] \ .
\end{eqnarray}
We further find that the contributions (without the
color-factors) from the diagrams (c) and (d) are opposite to each
other. Thus, their total contribution will be the difference on
the color-factor, which is $\frac{1}{N_c^2}\frac{if_{abc}}{2}{\rm
Tr}[T^aT^bT^c]$. That means the contributions from all these four
diagrams can be grouped into two terms with different color
factors: one with $C_F\times \frac{N_c^2-1}{4N_c^2}$, and one
with $\frac{1}{N_c^2}\frac{if_{abc}}{2}{\rm Tr}[T^aT^bT^c]$. The
latter one vanishes in the leading order of $P_{hT}/P_\perp$ after
we sum all diagrams contributions, and thus we are left with the
first color-factor contribution.

After summing over all diagrams' contribution, the spin-dependent
differential cross section coming from the Collins effect will be
\begin{eqnarray}
\frac{d\sigma(S_\perp)}{d{\cal
P.S.}}&=&\frac{\alpha_s^2}{\pi}\frac{N_c^2-1}{4N_c^2}\frac{4\hat
s\hat u}{-\hat t^2} {\epsilon^{\alpha\beta}S_\perp^\alpha}
\left[g^{\beta\beta'}-\frac{P_B^{\beta'}}{P_B\cdot
P_J}P_{J}^\beta\right]
\frac{g^2}{(2\pi)^3}C_Fg_s^2\int\frac{dq^+dq^-d^2q_T}{(2\pi)^4}
\nonumber\\
&&\times \left(q^+P_{hT}^{\beta'}-z_hq_T^{\beta'}\right)
\frac{1}{(k'-q)^2}\delta(q^2) \delta\left((P_h+k'-q)^2-M^2\right)
\ . \label{xs}
\end{eqnarray}
where $g_s$ is the strong coupling.
From the above result, we find a clear separation of the short
distance physics at the scale $P_\perp$ and long distance
physics at the scale $P_{hT}$. The short distance part is just
the hard factor $H_{qq'\to qq'}^{\rm Collins}$ for the
spin-dependent cross section, which can be calculated from the
partonic process with both initial and final state quarks
transversely polarized \cite{Stratmann:1992gu}, as we show in the
left panel of Fig.~5. The long distance part of the above result
can be factorized into the Collins fragmentation function
calculated from the right panel of Fig.~5. In this part, because
the $q_T^{\beta'}$ integral is proportional to $P_{hT}^{\beta'}$,
we can combine the two terms in the integral into one expression
contained in the Collins function. Therefore, the spin-dependent cross
section Eq.~(\ref{xs}) can be re-written as
\begin{eqnarray}
\frac{d\sigma(S_\perp)}{d{\cal
P.S.}}&=&\frac{\alpha_s^2}{\pi}\frac{N_c^2-1}{4N_c^2}\frac{4\hat
s\hat u}{-\hat t^2} {\epsilon^{\alpha\beta}S_\perp^\alpha}
\left[P_{hT}^\beta-\frac{P_B\cdot P_{hT}}{P_B\cdot
P_J}P_{J}^\beta\right] \frac{\delta \hat q(z_h,P_{hT})}{M_h} \ ,
\end{eqnarray}
where the Collins function $\delta \hat q$ is calculated from the
Feynman diagram in the right panel of Fig.~5,
\begin{eqnarray}
\delta\hat q (z_h,P_{hT})&=&\frac{M_h}{P_{hT}^\alpha}
\frac{g^2}{(2\pi)^3}g_s^2C_F\int\frac{dq^+dq^-d^2q_T}{(2\pi)^4}
\left(q^+P_{hT}^{\alpha}-z_hq_T^{\alpha}\right)\nonumber\\
&&\times  \frac{1}{(k'-q)^2-M^2}\delta(q^2)
\delta\left((P_h+k'-q)^2-M^2\right) \ ,
\end{eqnarray}
where the index $\alpha$ is not understood as a sum.
This final result demonstrates that we do have a factorization
for the spin-dependent cross section into the hard factor
$H_{qq'\to qq'}^{\rm Collins}$ times the Collins fragmentation
function, and the Collins function is the same as that in
$e^+e^-$ and SIDIS processes \cite{Amrath:2005gv}.

Therefore, by using the Ward identity at this particular order,
the final results for all the diagrams of Fig.~4 will sum up
together into a factorized form as shown in Fig.~5, where the
cross section is written as the hard partonic cross section for
$q(S_\perp)q'\to q(s_\perp)q'$ subprocess multiplied by a Collins
fragmentation function. The exchanged gluon in Fig.~4 is now
attaching to a gauge link from the fragmentation function
definition \cite{ColSop81p} as shown in the right panel of Fig.~5.

\begin{figure}[t]
\begin{center}
\includegraphics[width=10cm]{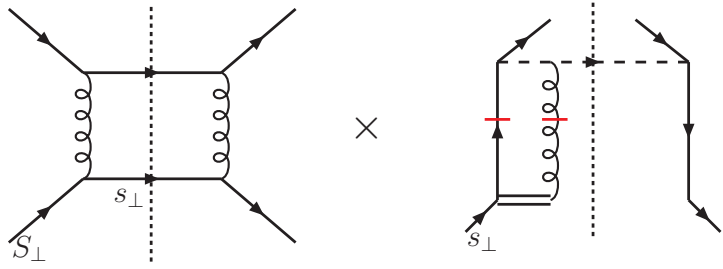}
\end{center}
\vskip -0.4cm \caption{\it Factorize the contributions from
Fig.~\ref{fig2} into the hard partonic cross section multiplied by
the universal Collins fragmentation function. The short bars
indicate the pole contribution to the Collins function.}
\label{fig3}
\end{figure}

The key steps in the above derivation are the eikonal
approximation and the Ward identity. The eikonal approximation is
valid when we calculate the leading power contributions in the
limit of $P_{hT}\ll k_\perp$. The Ward identity ensure that when
we sum up the diagrams with all possible gluon attachments we
shall get the eikonal propagator from the gauge link in the
definition of the fragmentation function. The most important
point to apply the Ward identity in the above analysis is that the
eikonal propagator does not contribute to the phase needed to
generate a nonzero SSA. This is what we have shown for the Collins
asymmetry in the above calculations, and the reason, as we mentioned above, is due to
the momentum flow of the exchanged gluon and the kinematic
constraints in the fragmentation process. We will show in the next
section, that for the two-gluon exchange diagrams the eikonal
propagators do not contribute to the phase for the nonzero Collins SSA
in this process. Therefore, we conjecture that the above conclusions are valid to
higher order contributions too.

This argument can not apply to the SSA associated with the parton
distributions, where the eikonal propagator does contribute to
the phase to generate a nonzero SSA. That is the reason we
have sign differences for the Sivers functions in SIDIS
and Drell-Yan processes.

\section{Two-gluon exchange contributions}

As we discussed in the last section, to demonstrate the
universality of the Collins function, we have to apply the Ward
identity to sum up all gluon exchange contributions into the
gauge link from the definition of the fragmentation function. In
order to use this argument, the eikonal propagator should not
contribute to the phase needed to generate nonzero SSA associated
with the Collins effects. This has been explicitly demonstrated
in the last section for the one-gluon exchange contribution.
In this section, we will extend the discussions to
the two-gluon exchange contributions. Especially, we will show
that these eikonal propagators do not contribute to the phase for
the SSAs. The reason, again, is due to the time-like feature and
the momentum flow in the fragmentation process.

\begin{figure}[t]
\begin{center}
\includegraphics[width=14cm]{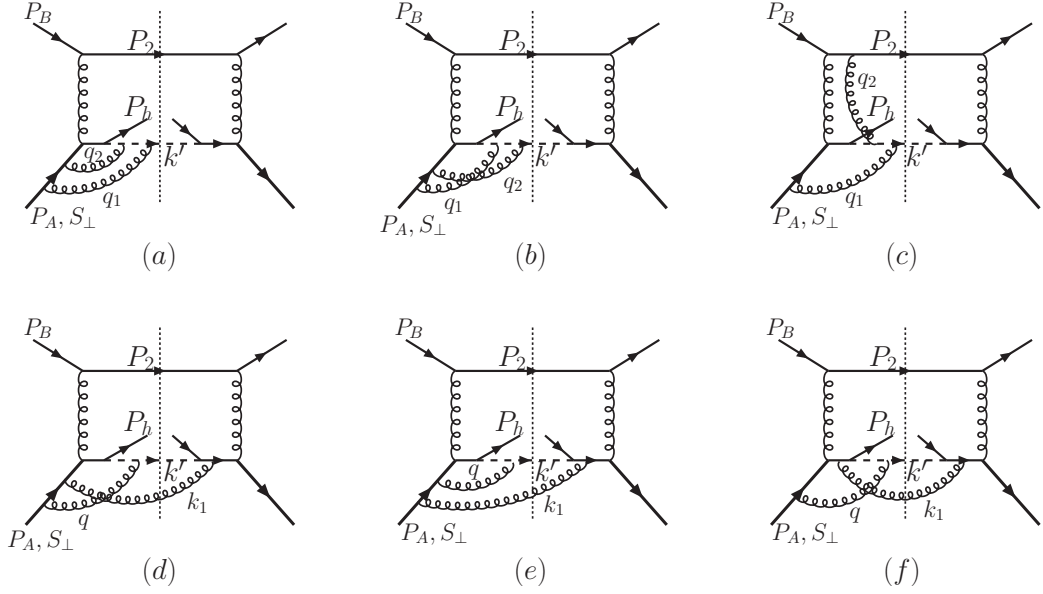}
\end{center}
\vskip -0.4cm \caption{\it Example diagrams for two-gluon exchange contributions (a,b,c); and
one real gluon radiation contributions (d,e,f).}
\label{fig4}
\end{figure}

We will focus our discussions on some representative diagrams from
the two-gluon exchange contributions. All other diagrams will
follow accordingly. We show these diagrams in Figs.~6(a,b,c). The
contribution from Fig.~6(a) will depend on the following integral
of the exchange gluons' momenta $q_1$ and $q_2$,
\begin{eqnarray}
&&\int\frac{d^4q_1}{(2\pi)^4}\frac{d^4q_2}{(2\pi)^4} {\cal
M}(q_1,q_2)
\frac{1}{(P_A-q_1)^2+i\epsilon}\frac{1}{(P_A-q_1-q_2)^2+i\epsilon}
\frac{1}{(k'-q_1)^2+i\epsilon}\nonumber\\
&&~~\frac{1}{(k'-q_1-q_2)^2+i\epsilon}
\frac{1}{(k-q_1-q_2)^2+i\epsilon}
\frac{1}{q_1^2+i\epsilon}\frac{1}{q_2^2+i\epsilon} \ ,
\end{eqnarray}
where $k=P_1=k'+P_h$ is the fragmenting quark's momentum and
${\cal M}(q_1,q_2)$ represents the numerators depending $q_1$ and
$q_2$, especially their transverse momentum components. Following
the arguments used in the last section, the first two propagators
in the above expression can be further simplified by using the
eikonal approximation, and then we will obtain the following
expression
\begin{eqnarray}
&&\int\frac{dq_1^-dq_1^+}{(2\pi)^2}\frac{dq_2^-dq_2^+}{(2\pi)^2}
\frac{1}{-q_1^++i\epsilon}\frac{1}{-q_1^+-q_2^++i\epsilon}
\frac{1}{(k'-q_1)^2+i\epsilon}\frac{1}{(k'-q_1-q_2)^2+i\epsilon}
\nonumber\\
&&~~
\frac{1}{(k-q_1-q_2)^2+i\epsilon}\frac{1}{q_1^2+i\epsilon}\frac{1}{q_2^2+i\epsilon}
\ ,\label{eq1}
\end{eqnarray}
where $q_i^\pm$ follow the definitions in the last section. The
normalization of the above integral has been changed for
convenience. This normalization is not relevant for our
discussions, because we want to show that the eikonal propagators
do not contribute to the phase needed for a nonzero SSA, not the
actual contribution from this diagram. We will show if we take
pole contributions from these two eikonal propagators, the final
integral will vanish. Because of the existence of two eikonal
propagators, the analysis will be more complicated than that in
the last section. We discuss their contributions separately.
\begin{enumerate}
\item pole contribution from $\frac{1}{-q_1^+-q_2^++i\epsilon}$.

If we take pole of this eikonal propagator, $q_1^+$ and $q_2^+$
will be constrained: $q_1^++q_2^+=0$, and the integral of
(\ref{eq1}) will become,
\begin{eqnarray}
&&\int
\frac{dq_1^+dq_2^+}{2\pi}\frac{\delta(q_1^++q_2^+)}{q_1^+}\int
\frac{dq_1^-dq_2^-}{(2\pi)^2} \frac{1}{-2(k^{\prime
+}-q_1^+)q_1^-+\Delta_1+i\epsilon} \frac{1}{-2k^{\prime
+}(q_1^-+q_2^-)+\Delta_2+i\epsilon}\nonumber\\
&&~~\frac{1}{-2k^+(q_1^-+q_2^-)+\Delta_3+i\epsilon}
\frac{1}{2q_1^+q_1^-+\Delta_4+i\epsilon}\frac{1}{2q_2^+q_2^-+\Delta_5+i\epsilon}
\ , \label{eq3}
\end{eqnarray}
where $\Delta_i$ are some quantities depending on the transverse
momenta of $q_i$, $k'$ and $P_h$. The following analysis does not
depend on the details of these numbers. In deriving the above
equation, we have used the constraint of $q_1^++q_2^+=0$ to
simplify the expression. We further notice that $k^{\prime +}>0$
and $k^+>0$. Thus, the poles of the second and third factor in the
integral of $q_1^-$ and $q_2^-$ are both in the upper half plane.
If $q_1^+>0$, which means that $q_2^+<0$, the pole of the fifth
factor will be also in the upper half plane of $q_2^-$.
Therefore, the poles of the three factors (the second, third and
fifth) depending on $q_2^-$ are all in the upper half plane of
$q_2^-$, and the integral over $q_2^-$ will vanish, and so will
the above integral. Similarly, if $q_1^+<0$, the pole of the
fourth factor will be in the upper half plane of $q_1^-$.
Meanwhile, we will also have $k^{\prime +}-q_1^+>0$, and the pole
of the first factor will be in the upper half plane too.
Therefore, the poles of the four factors (the first, second,
third and fourth) depending on $q_1^-$ are all in the upper half
plane of $q_1^-$. The integral over $q_1^-$ will vanish, and so
will the above expression.  In conclusion, in any case of
$q_1^+>0$ or $q_1^+<0$, the above integral vanishes, and we do
not have contribution from the pole of
$\frac{1}{-q_1^+-q_2^++i\epsilon}$.

\item pole contribution from $\frac{1}{-q_1^++i\epsilon}$

Because $q_1^+=0$, we can simplify the integral of (\ref{eq1}) as
follows,
\begin{eqnarray}
&&\int\frac{dq_1^+dq_2^+}{2\pi}\frac{\delta(q_1^+)}{q_2^+}\int
\frac{dq_1^-dq_2^-}{(2\pi)^2} \frac{1}{-2k^{\prime
+}q_1^-+\Delta_1+i\epsilon}
\frac{1}{-2(k^{\prime+}-q_2^+)(q_1^-+q_2^-)+\Delta_2+i\epsilon}\nonumber\\
&&~~ \frac{1}{-2(k^{+}-q_2^+)(q_1^-+q_2^-)+\Delta_3+i\epsilon}
\frac{1}{2q_2^+q_2^-+\Delta_5+i\epsilon} \ . \label{eq4}
\end{eqnarray}
Again, the normalization has been changed for convenience. Because
$k^+>k^{\prime+}$, we will analyze the contributions of the above
equation by classifying the different regions of $q_2^+$: (a)
$q_2^+>k^+$; (b) $q_2^+<k^{\prime +}$; (c)
$k^{\prime+}<q_2^+<k^+$. In the region of (a), we will have
$k^+-q_2^+<0$ and $k^{\prime+}-q_2^+<0$. Therefore, the poles of
the three factors (the second, third and fourth) are all in the
lower half plane of $q_2^-$, and the integral over $q_2^-$
vanishes. In the region of (b), we have $k^{\prime+}>0$,
$k^{\prime+}-q_2^+>0$ and $k^+-q_2^+>0$. Thus, the poles of the
three factors (the first, second and third) depending on $q_1^-$
are all in the upper half plane, and the integral over $q_1^-$
vanishes. In the region of (c), we have $q_2^+>0$,
$k^{\prime+}-q_2^+<0$ and $k^+-q_2^+>0$. Therefore, the $q_2^-$
integral will pick up the pole of the third factor, which
actually determines the value of $q_1^-+q_2^-$. After
substituting this back into the equation, we will find the second
factor does not depend on $q_1^-$ any more. The only dependence
comes from the first factor. Obviously, this integral over
$q_1^-$ will vanish. In conclusion, in any case of (a,b,c), the
above integral vanishes, and there is no contribution from the
pole of $\frac{1}{-q_1^++i\epsilon}$.

\end{enumerate}

In summary, there is no contribution to the SSA from the pole of
the eikonal propagators in the diagram of Fig.~6(a). Similarly,
the contribution from Fig.~6(b) depends on the following integral,
\begin{eqnarray}
&&\int\frac{dq_1^-dq_1^+}{(2\pi)^2}\frac{dq_2^-dq_2^+}{(2\pi)^2}
\frac{1}{-q_1^++i\epsilon}\frac{1}{-q_1^+-q_2^++i\epsilon}
\frac{1}{(k'-q_2)^2+i\epsilon}\frac{1}{(k'-q_1-q_2)^2+i\epsilon}\nonumber\\
&&~~ \frac{1}{(k-q_1-q_2)^2+i\epsilon}
\frac{1}{q_1^2+i\epsilon}\frac{1}{q_2^2+i\epsilon} \ ,\label{eq2}
\end{eqnarray}
where we have made the eikonal approximations for the two
propagators along the incident quark line $P_A$. Comparing with
Eq.~(\ref{eq1}), we find the only difference is the third factor
$q_1\to q_2$. Again, we can show that none of the two eikonal
propagators will contribute to the phase needed for a nonzero SSA.
We will discuss their contributions separately.

\begin{enumerate}

\item pole contribution from $\frac{1}{-q_1^+-q_2^++i\epsilon}$.

After taking this pole, the integral of (\ref{eq2}) will become,
\begin{eqnarray}
&&\int
\frac{dq_1^+dq_2^+}{2\pi}\frac{\delta(q_1^++q_2^+)}{q_1^+}\int
\frac{dq_1^-dq_2^-}{(2\pi)^2} \frac{1}{-2(k^{\prime
+}-q_2^+)q_2^-+\Delta_1+i\epsilon} \frac{1}{-2k^{\prime
+}(q_1^-+q_2^-)+\Delta_2+i\epsilon}\nonumber\\
&&~~\frac{1}{-2k^+(q_1^-+q_2^-)+\Delta_3+i\epsilon}
\frac{1}{2q_1^+q_1^-+\Delta_4+i\epsilon}\frac{1}{2q_2^+q_2^-+\Delta_5+i\epsilon}
\ ,
\end{eqnarray}
which is the same as Eq.~(\ref{eq3}) if we interchange $q_1^\pm$
and $q_2^\pm$. Thus, the above integral will vanish by the same
arguments we have used for Eq.~(\ref{eq3}).

\item pole contribution from $\frac{1}{-q_1^++i\epsilon}$

Because $q_1^+=0$, we can simplify the integral of (\ref{eq2}) as
follows,
\begin{eqnarray}
&&\int\frac{dq_1^+dq_2^+}{2\pi}\frac{\delta(q_1^+)}{q_2^+}\int
\frac{dq_1^-dq_2^-}{(2\pi)^2} \frac{1}{-2(k^{\prime
+}-q_2^+)q_2^-+\Delta_1+i\epsilon}
\frac{1}{-2(k^{\prime+}-q_2^+)(q_1^-+q_2^-)+\Delta_2+i\epsilon}\nonumber\\
&&~~ \frac{1}{-2(k^{+}-q_2^+)(q_1^-+q_2^-)+\Delta_3+i\epsilon}
\frac{1}{2q_2^+q_2^-+\Delta_5+i\epsilon} \ .\label{eq6}
\end{eqnarray}
Again, we classify three different regions of $q_2^+$ in the
above equation: (a) $q_2^+>k^+$; (b) $q_2^+<k^{\prime +}$; (c)
$k^{\prime+}<q_2^+<k^+$. The contributions from (a) and (b)
regions vanish by the same reasons as we have shown for
Eq.~(\ref{eq4}) in the above. In region (c), we have
$k^{\prime+}-q_2^+<0$ and $k^+-q_2^+>0$. Therefore, the $q_1^-$
integral will pick up the pole of the third factor, which again
actually determines the value of $q_1^-+q_2^-$. After
substituting this back into the equation, we will find the second
factor does not depend on $q_2^-$ any more. The only dependence
comes from the first and last factors. Obviously, this integral
over $q_2^-$ vanishes because $k^{\prime+}-q_2^+<0$ and $q_2^+>0$,
and the poles of these two factors are both in the lower half
plane. In conclusion, in any case of (a,b,c), the above integral
vanishes, and there is no contribution from the pole of
$\frac{1}{-q_1^++i\epsilon}$.
\end{enumerate}
Similarly, the contribution from Fig.~6(c) will depend on the
following integral,
\begin{eqnarray}
&&\int\frac{dq_1^-dq_1^+}{(2\pi)^2}\frac{dq_2^-dq_2^+}{(2\pi)^2}
\frac{1}{-q_1^++i\epsilon}\frac{1}{q_2^++i\epsilon}
\frac{1}{(k'-q_1)^2+i\epsilon}\frac{1}{(k'-q_1-q_2)^2+i\epsilon}
\nonumber\\
&&~~
\frac{1}{(k-q_1-q_2)^2+i\epsilon}\frac{1}{q_1^2+i\epsilon}\frac{1}{q_2^2+i\epsilon}
\ ,\label{eq5}
\end{eqnarray}
where again we have made the eikonal approximations. There are
two eikonal propagators in the above equation, and as above we
will discuss their contributions separately.

\begin{enumerate}

\item the pole contribution from $\frac{1}{-q_1^++i\epsilon}$

After taking this pole, $q_1^+=0$, the above equation
Eq.~(\ref{eq5}) will reduce to Eq.~(\ref{eq4}). According to the
same arguments we used there, there will be no contributions from
this pole.

\item the pole contribution from $\frac{1}{q_2^++i\epsilon}$

This pole contribution means that $q_2^+=0$, and the integral of
Eq.~(\ref{eq5}) become
\begin{eqnarray}
&&\int\frac{dq_1^+dq_2^+}{2\pi}\frac{\delta(q_2^+)}{q_1^+}\int
\frac{dq_1^-dq_2^-}{(2\pi)^2} \frac{1}{-2(k^{\prime
+}-q_1^+)q_1^-+\Delta_1+i\epsilon}
\frac{1}{-2(k^{\prime+}-q_1^+)(q_1^-+q_2^-)+\Delta_2+i\epsilon}\nonumber\\
&&~~ \frac{1}{-2(k^{+}-q_1^+)(q_1^-+q_2^-)+\Delta_3+i\epsilon}
\frac{1}{2q_1^+q_1^-+\Delta_5+i\epsilon} \ ,\label{eq7}
\end{eqnarray}
which will be identical to Eq.~(\ref{eq6}) if we interchange
$q_1^\pm$ to $q_2^\pm$. Using the same arguments there, the above
integral vanishes.
\end{enumerate}

The above three examples are typical diagrams we encounter for
the two-gluon exchange contributions for this channel. All these
diagrams can be analyzed by a similar manner, and we will find that
the eikonal propagators do not contribute to the phase needed to
a nonzero SSA. Because of this fact, all these diagrams can be
summed together to form the contributions from the gauge link in
the fragmentation function, where the two gluons attach to the
gauge link similar to the diagram we have shown in Fig.~5. Since
there is no contributions from these eikonal propagators, the
Collins function calculated from these diagrams will be the same
as that in $e^+e^-$ and SIDIS processes, and the universality
preserved.

We have also drawn some other diagrams at this order in
Fig.~6(d,e,f), which contribute to a real gluon radiation in
addition to the gluon exchange. The analysis of these diagrams
also show that we do not get contribution from the pole of the
eikonal propagators. For example, the contribution from Fig.~6(d)
depends on
\begin{eqnarray}
\int\frac{d^4q}{(2\pi)^4}\frac{1}{-q^++i\epsilon}\frac{1}{-q^+-k_1^++i\epsilon}
\frac{1}{(k'-q)^2+i\epsilon}
\frac{1}{(k-q-k_1)^2+i\epsilon}\frac{1}{q^2+i\epsilon} \ ,
\end{eqnarray}
where $k_1$ is the momentum for the radiated gluon. We have two
eikonal propagators in the above equation. However, none of them
contributes to the phase needed to a nonzero SSA.
\begin{enumerate}
\item the pole contribution from $\frac{1}{-q^++i\epsilon}$

This pole contribution means that $q^+=0$, and the integral of
$q^-$ will reduce to
\begin{equation}
\int
\frac{dq^-}{2\pi}\frac{1}{-2k^{\prime+}q^-+\Delta_1+i\epsilon}
\frac{1}{-2(k^+-k_1^+)q^-+\Delta_2+i\epsilon} \ .
\end{equation}
Because $k^+=k_1^++k^{\prime+}+P_h^+>k_1^+$ and $k^{\prime +}>0$,
the poles of the above two factors are both in the lower half
plane of $q^-$, and the integral vanishes.

\item the pole contribution from $\frac{1}{-q^+-k_1^++i\epsilon}$

After taking this pole, we will have the following $q^-$ integral
\begin{equation}
\int
\frac{dq^-}{2\pi}\frac{1}{-2(k^{\prime+}-q^+)q^-+\Delta_1+i\epsilon}
\frac{1}{-2k^+q^-+\Delta_2+i\epsilon}
\frac{1}{2q^+q^-+\Delta_3+i\epsilon}\ .
\end{equation}
Because the pole constrains that $q^+=-k_1^+<0$ and
$k^{\prime+}-q^+>0$, the poles of the above three factors are all
in the lower half plane. The integral over $q^-$ vanishes.
\end{enumerate}
In summary, there is no contribution from the pole of the eikonal
propagators in the diagram of Fig.~6(d).

The contribution from Fig.~6(e) will depend on the following
integral,
\begin{eqnarray}
\int\frac{d^4q}{(2\pi)^4}\frac{1}{-k_1^++i\epsilon}\frac{1}{-q^+-k_1^++i\epsilon}
\frac{1}{(k'-q)^2+i\epsilon}
\frac{1}{(k-q-k_1)^2+i\epsilon}\frac{1}{q^2+i\epsilon} \ .
\end{eqnarray}
Because $k_1^+>0$, we only have one possible pole contribution
from the eikonal propagator $1/(-q^+-k_1^++i\epsilon)$, which
vanishes by the same reason as above for diagram Fig.~6(d).
Similarly, if the gluon with momentum $q$ attaches to the
radiated gluon instead of the incident quark line with momentum
$P_A$ (we did not show this diagram in Fig.~6), the contribution
vanishes by the same reason.

The contribution from Fig.~6(f) depends on the following integral
\begin{eqnarray}
\int\frac{d^4q}{(2\pi)^4}\frac{1}{-q^++i\epsilon}
\frac{1}{(k'-q)^2+i\epsilon}
\frac{1}{(k-q-k_1)^2+i\epsilon}\frac{1}{(k-q)^2+i\epsilon}
\frac{1}{q^2+i\epsilon} \ ,
\end{eqnarray}
after eikonal approximation. If we take the pole contribution
from the eikonal propagator, the above integral will reduce to
\begin{equation}
\int\frac{dq^-}{2\pi}\frac{1}{-2k^{\prime+}q^-+\Delta_1+i\epsilon}
\frac{1}{-2(k^+-k_1^+)q^-+\Delta_2+i\epsilon}
\frac{1}{-2k^+q^-+\Delta_3+i\epsilon} \ .
\end{equation}
Again, because $k^+>k_1^+$, the poles of the above three factors
are all in the lower half plane, and the integral over $q^-$
vanishes. Thus, there is no contribution from the pole of the
eikonal propagator for this diagram.

In summary, for the gluon radiation diagrams, there is no
contributions from the pole of the eikonal propagators. Because
of this fact, we can use Ward identity to sum all these
diagrams together to form the gauge link contribution from the
fragmentation function, similar to the diagram in Fig.~5 with an
additional gluon radiation.

Concluding the analysis of the two-gluon exchange
diagrams in Fig.~6, the eikonal propagators do not contribute to
the phase needed for the nonzero SSA associated with the Collins
effect. Therefore, we can apply the Ward identity at this order
to sum all these diagrams plus other similar ones. This sum will
lead to the gauge link contribution from the fragmentation
function definition, and the fragmentation function will be the same
as that in $e^+e^-$ and SIDIS processes.

\section{Summary and Discussions}

In this paper, we have shown that the Collins function in hadron
production in single-transverse-spin polarized $pp$ scattering is
the same as that in $e^+e^-$ and SIDIS processes.
This universality is a general and model-independent
observation, and depends on the fact that the eikonal
propagators do not contribute to the phase needed for a nonzero
SSA. We have demonstrated this by explicit calculations for
one-gluon exchange diagrams which corresponds to one eikonal
propagator in the amplitudes, and two-gluon exchange diagrams
which correspond to two eikonal propagators.
Although our calculations were based on a model \cite{Col93,{Manohar:1983md}},
the analysis and arguments are quite general. The results,
as we emphasized, depend on the momentum flow and kinematic
constraints in the fragmentation process.

This observation is very different from the SSAs associated
with the parton distributions, where the eikonal propagators from
the gauge link in the parton distribution definition play very
important role. It is the pole of these eikonal propagators
contribute to the phase needed for a nonzero SSA associated with
the naive-time-reversal-odd parton distributions, which also
predicts a sign difference for the quark Sivers function between
the SIDIS and Drell-Yan processes. More complicated results have
been found for the SSAs in the hadronic dijet-correlation~\cite{gl,QVY},
where a normal TMD factorization
breaks down \cite{CQ}. The reason is that the eikonal propagators from the
initial and final state interactions in dijet-correlation process
do contribute poles in the cross section~\cite{QVY,CQ}. Because of this, the
Ward identity is not applicable, and the standard TMD factorization
breaks down, although a modified factorization may be valid if we
modify the definition of the TMD parton distributions to take
into account all the initial and final state interaction effects~\cite{gl}.
In the fragmentation process, as we discussed in our paper, the eikonal propagators do not
contribute to an imaginary part, and the Ward identity is applicable. We have shown this
in our explicit calculations including one-gluon and two-gluon exchange contributions.

There has been discussion about the twist-three quark-gluon
correlation contribution in the fragmentation function, especially for the Collins effects
~\cite{BoeMulPij03,{Kanazawa:2000hz}}. It will be interesting to further understand
these contributions following the analysis in this paper, and discuss the universality
issues in a more general ground \cite{Metz:2002iz,{Collins:2004nx},{collins-s}}.

We thank J.~Collins, L.~Gamberg, R.~Jaffe, X.~Ji, J.~Qiu, and W.~Vogelsang for useful
discussions. Especially, we thank A. Metz for his valuable comments and many
useful discussions. This work was supported in part by the U.S. Department of Energy
under contract DE-AC02-05CH11231. We are grateful to RIKEN,
Brookhaven National Laboratory and the U.S. Department of Energy
(contract number DE-AC02-98CH10886) for providing the facilities
essential for the completion of this work.

\end{document}